\documentclass[12pt]{article}

\usepackage{amsmath,amssymb}

\begin{document}

\title{\bf $\hbar$-expansion for the Schr\"odinger equation with a position-dependent mass}
\date{}

\author{D.~A.~Kulikov\thanks{kulikov\_d\_a@yahoo.com, kulikov@dsu.dp.ua}
\\
\em{\small Theoretical Physics Department, FFEKS, Dniepropetrovsk
National University }\\
\em{\small 72 Gagarin avenue, Dniepropetrovsk 49010, Ukraine}\\
$\phantom{line}$\\
V.~M.~Shapoval\thanks{shapoval@bitp.kiev.ua}\\
\em{\small Bogolyubov Institute for Theoretical Physics}\\
\em{\small 14-b Metrolohichna str., Kiev 03680, Ukraine}\\
}

\maketitle

\begin{abstract}

A recursion technique of obtaining the asymptotical expansions for
the bound-state energy eigenvalues of the radial Schr\"odinger
equation with a position-dependent mass is presented. As an
example of the application we calculate the energy eigenvalues for
the Coulomb potential in the presence of position-dependent mass
and we derive the inequalities regulating the shifts of the energy
levels from their constant-mass positions.


PACS number(s): 03.65.-w, 03.65.Ge, 03.65.Sq

\end{abstract}

\section*{Introduction}
\label{part1}

The notion of effective mass, which was introduced more than 50
years ago to describe the motion of electrons in crystals
\cite{slater}, has now a broader scope. It is
employed in nuclear physics \cite{serot}, theory of quantum
liquids \cite{quantliquid}, metallic clusters  \cite{clusters} and
semiconductor heterostructures: quantum wells, wires and dots
\cite{bastard}. Attempts to accommodate spatial inhomogeneity of
multi-layer nanostructures have invoked  treatment of the
effective mass as a position-dependent quantity
\cite{young,herling,borovitskaya}.

In order to calculate the bound-state spectrum of the
Schr\"odinger equation with a position-dependent mass (PDM),
various iterative \cite{li,voss}, variational
\cite{muljarov,tkach07} and perturbative \cite{tkach} schemes were
proposed. Usually, these schemes provide solutions in numerical
form. On the other hand, there exist methods of deriving exact
analytical solutions to the PDM Schr\"odinger equation for some
potentials such as the Coulomb, oscillator and Morse potentials
\cite{alhaidari,tkachuk,tkachuk2005}. Having a solution in
analytical form greatly simplifies investigation of the structure
of energy levels. Thus an important task is the development of
approximate analytical methods to solve the bound-state problem
for the PDM Schr\"odinger equation.

For the constant-mass Schr\"odinger equation with spherical
symmetry, the $1/N$-expansion method has proved to be efficient.
The name of $1/N$-expansion refers to the group of related
approaches \cite{mlodinow,imbo,weinberg,tutik,high1N} that start
with describing classical motion of a particle located at the
bottom of the effective-potential well and then consider quantum
corrections. The versions of this method differ in choosing a
formal expansion parameter. Namely, going to the classical limit,
one takes for the small parameter the inverse number of spatial
dimensions, the inverse principal quantum number, or the Planck
constant. The last choice, the so-called
$\hbar$-expansion~\cite{tutik}, seems to be the most natural one
and, besides, allows to formalize calculations in a simple
recursion procedure.

The purpose of this work is to extend the $\hbar$-expansion
technique to the PDM Schr\"odinger equation. We develop the
procedure for calculating the bound-state energy eigenvalues and
then apply it to investigate the influence of PDM on the Coulomb
potential spectrum.

\section{$\hbar$-expansion technique}
\label{part2}

The PDM Hamiltonian of the particle is given by~\cite{roos}
\begin{equation}\label{eq21}
\hat{H}= \hat{T} + V (\textbf{r}), \quad \hat{T} = -
\frac{\hbar^2}{4} \left[ m (\textbf{r})^{\alpha} \boldsymbol\nabla
m(\textbf{r})^{\beta} \boldsymbol\nabla m(\textbf{r})^{\gamma} +
m(\textbf{r})^{\gamma} \boldsymbol\nabla m(\textbf{r})^{\beta}
\boldsymbol\nabla m(\textbf{r})^{\alpha} \right],
\end{equation}
where $m(\textbf{r})$ is the PDM,  $V(\textbf{r})$ is the
interaction potential, $\alpha, \beta, \gamma$  are the ambiguity
parameters ($\alpha + \beta + \gamma = -1$).

Let us consider the spherically symmetric case in which the
stationary Schr\"odinger equation with the Hamiltonian
(\ref{eq21}) is reduced to the radial equation
\begin{eqnarray}\label{eq1}
&& \left\{ \frac{d^2}{dr^2} + \frac{m'(r)}{m(r)} \left( \frac{1}{r} -
\frac{d}{dr} \right) - \frac{l (l + 1)}{ r^2} - \frac{2 m(r)}{\hbar^2}
(U(r) - E)  \right\}\psi (r) = 0 , \nonumber\\
&& U(r) = V(r) - \frac{\hbar^2}2 \left[ \frac{\alpha +
\gamma}{2}\left( \frac{m''(r)}{m^2(r)}+ \frac{2m'(r)}{m^2(r)r}
\right) - (\alpha \gamma + \alpha + \gamma) \frac{{m'}^2
(r)}{m^3(r)} \right] ,
\end{eqnarray}
with $l$ being the orbital quantum number and primes denoting
derivatives with respect to $r=|\textbf{r}|$.

Assuming $V(r)$ and $m(r)$ are analytical functions such that
Eq.~(\ref{eq1}) possesses bound states, we intend to find the
discrete energy spectrum. To that end, we generalize the
$\hbar$-expansion technique~\cite{tutik} to the PDM case.

First we recast Eq.~(\ref{eq1}) into the Riccati equation for the
logarithmic derivative of the wave function. Upon substituting
$\psi (r) = \chi (r) \sqrt{m (r)}$ to eliminate the term with
$d\psi (r)/dr$ and then putting
 $C(r)=\hbar \chi ' (r)/\chi (r)$, we get
\begin{eqnarray}\label{eq2}
&&\hbar C'(r) + C^2 (r) =\frac{\hbar^2 l (l+1)}{r^2} + \left( \frac34 + \alpha \gamma + \alpha + \gamma \right) \hbar^2 Q^2 (r) - \frac{\hbar^2}{r} Q(r) - \nonumber \\
&& - \frac{1 + \alpha + \gamma}{2}  \hbar^2 P(r) + 2 m(r) (V (r) -
E)
\end{eqnarray}
where we denoted $Q(r) = m'(r)/m(r)$, $P(r) =
(m''(r)+2m'(r)/r)/m(r)$.

We try to obtain the solution to the Riccati equation with series
expansions in the Planck constant $\hbar$
\begin{equation}\label{eq3}
C(r) = \sum_{k = 0}^{\infty} C_{k}(r) \hbar^{k} , \qquad E =
\sum_{k = 0}^{\infty} E_{k} \hbar^{k}.
\end{equation}

In view of the freedom in defining the classical limit, which is
formally the limit of  $\hbar \to 0$, the centrifugal contribution
can be rewritten as follows
\begin{equation}\label{eq4}
\hbar^2 l (l+1) = \Lambda^2 + \hbar A \Lambda + \hbar^2 B
\end{equation}
that enables to consider different modifications of the method
depending on values of parameters  $A$ and $B$ \cite{tutik}. A
particular choice of these values amounts to fixing a zeroth
approximation. In practice, one first specifies $V(r)$ and $m(r)$
and then choose  $A$ and $B$ in such a way that higher-order
corrections in the $\hbar$-expansions for eigenenergies
(\ref{eq3}) become as small as possible. Note that one may achieve
yet faster convergence of this series expansion by means of the
renormalization: rewriting $V(r)$ or $m(r)$ as the
$\hbar$-expansion and properly adjusting its coefficients. Such
generalizations of the method lie however out of the scope of the
present article.

Substituting the expansions (\ref{eq3}) into the Riccati equation
(\ref{eq2}) and equating to zero coefficients of successive powers
of $\hbar$, we obtain the system of equations
\begin{eqnarray}\label{eq5}
&&C_0^2 = 2 m (r)(V(r) - E_0) + \frac{\Lambda^2}{r^2} ,\nonumber \\
&&C_0' + 2 C_0 C_1 = - 2 m (r) E_1 + \gamma_1 {\left( \frac{r_0}r \right)}^2 , \nonumber \\
&&C_1' + 2 C_0 C_2 + C_1^2 = - 2 m (r) E_2 + \gamma_2 {\left( \frac{r_0}{r} \right)}^2 + F (r)  ,\nonumber\\
&& \ldots \nonumber \\
&&C_{k-1}' + \sum_{i = 0}^{k} C_i C_{k - i} = - 2 m (r) E_{k} +
\gamma_k {\left( \frac{r_0}r \right)}^2  , \, k > 2 .
\end{eqnarray}
Here $\gamma_1 = {A \Lambda}/{r_0^2}$, $\gamma_2 = {B}/{r_0^2}$,
$\gamma_3 = \gamma_4 = \ldots = 0 $,
\begin{equation}\label{eq6}
F (r) = \left(\frac34 + \alpha \gamma +\alpha + \gamma \right)
Q^2(r) - \frac1r Q(r) - \frac{1 + \alpha + \gamma}2 P(r) .
\end{equation}

In order to take into account the nodes of the wave function for
radially excited states, we employ the Zwaan-Dunham quantization
condition \cite{zwaan,dunham} which expresses the principle of
argument in complex analysis. For the discrete spectrum, solutions
to Eq.~(\ref{eq1}) are real on the real axis and possess a finite
number of simple nodes on it. The number of these nodes is just
the radial quantum number $n_r$. Then the integration of the
logarithmic derivative $C(r)$ along the contour that encloses only
the above nodes yields
\begin{equation}\label{eq7}
\frac1{2 \pi i} \oint C(r) dr = n_r \hbar, \quad n_r = 0, 1, 2, \ldots
\end{equation}

This condition must be supplemented with a rule of achieving the
classical limit $\hbar \to 0$ for the radial and orbital quantum
numbers which are the specific quantum notions. We choose this
rule as follows
\begin{equation}\label{eq8}
\hbar \to 0, ~~ n_r = const, ~~ l \to \infty, ~~ \hbar n_r \to 0, ~~ \hbar l = const .
\end{equation}
Physically, it means that in the classical limit the particle is
located in the minimum of the effective potential $V(r)
+\Lambda^2/(2 m (r) r^2)$ and thus moves along the stable circular
orbit whose radius $r_0$ is a solution to the equation
\begin{equation}\label{eq9}
m (r_0) r_0^3 V' (r_0) = \Lambda^2 \left( 1 + \frac{m' (r_0) r_0}{2 m (r_0)} \right).
\end{equation}

Meanwhile, the energy in the zeroth approximation is given by
\begin{equation}\label{eq10}
E_0 = V(r_0) + \frac{\Lambda^ 2}{2 m (r_0) r_0^2}
\end{equation}
and the quantization rule transforms into
\begin{equation}\label{eq11}
\frac1{2 \pi i} \oint C_k(r) dr = n_r \delta_{k,1}.
\end{equation}

With the help of such rules we can develop an algebraic recursion
procedure for calculation of the $\hbar$-expansion corrections to
energies avoiding bulky formulas that arise in higher orders of
the standard Rayleigh-Schr\"odinger perturbation theory.

\section{Recursion
procedure for calculating energies} \label{part3}

To solve Eqs.~(\ref{eq5}), we expand the functions $V(r)$, $m(r)$,
$F(r)$, $P(r)$ and $Q(r)$ into Taylor series in powers of the new
variable, the deviation from the minimum of the effective
potential, $x = (r - r_0) / r_0$,
\begin{eqnarray}\label{eq110}
&& V(r)=\sum_{i=0}^{\infty}V_i x^i, \quad
m(r)=\sum_{i=0}^{\infty}m_i
x^i, \quad F(r)=\sum_{i=0}^{\infty}F_i x^i, \nonumber\\
&& P(r)=\sum_{i=0}^{\infty}P_i x^i,\quad
Q(r)=\sum_{i=0}^{\infty}Q_i x^i
\end{eqnarray}
where $V_i =r_0^i V^{(i)}(r_0)/i!$, $m_i =r_0^i m^{(i)}(r_0)/i!$,
whereas $F_i$, $P_i$ and $Q_i$ are expressed through $V_i$, $m_i$.

Consider the first equation of Eqs.~(\ref{eq5}) that determines
$C_0(r)$. Writing down the squared quantity as
\begin{equation}\label{eq13}
C_0^2(r)=\omega^2x^2 (1 + a_1 x + a_2 x^2 + \ldots)
\end{equation}
with
\begin{eqnarray}\label{eq14}
&& \omega = \sqrt{2 (m_0 V_2 + m_1 V_1) + \frac{2 m_0 V_1}{2 m_0 + m_1} (3 m_0 - m_2)}, \nonumber\\
&& a_i = \frac{2}{\omega^2} \left( \sum_{j = 0}^{i + 1} m_j
V_{i-j+2} + \frac{m_0^2 V_1}{2 m_0 + m_1} \left( (i + 3) (-1)^i -
\frac{m_{i + 2}}{m_0} \right) \right),
\end{eqnarray}
we find the coefficients of the Taylor series expansion for
$C_0(x)$
\begin{equation}\label{eq15}
C_0 (x) = - \omega x \sqrt{1 + a_1 x + a_2 x^2 + \ldots} = x
\sum_{i = 0}^{\infty} C_i^0 x^i ,
\end{equation}
\begin{equation}\label{eq150}
 C_0^0 = - \omega, \quad C_i^0 = \frac{1}{2 \omega}
(\sum_{j=1}^{i-1} C_j^0 C_{i-j}^0 - \omega^2 a_i).
\end{equation}
The "$-$"\,  sign in Eq.~(\ref{eq15}) is chosen so as to meet the
boundary conditions: $C_0(r)>0$ for $r<r_0$, $C_0(r)<0$ for
$r>r_0$, which follow from the quadratic integrability of the wave
function.

The function $C_0(x)$ has a simple zero at $x = 0$. Then, after
examining the system of equations  (\ref{eq5}), we conclude that
$C_k(x)$ has a pole of the order of $(2k-1)$ at this point and,
hence, can be represented by the Laurent series
\begin{equation}\label{eq16}
C_k (x) = x ^ {1-2 k} \sum_{i=0}^{\infty} C_i^k x^i .
\end{equation}

With definition of residues, this expansion permits us to express
the quantization conditions (\ref{eq11}) explicitly in terms of
the coefficients $C_i^k$
\begin{equation}\label{eq17}
C^k_{2k-2} = \frac{n_r}{r_0} \delta_{1, k} .
\end{equation}

Inserting the above expansions into Eqs.~(\ref{eq5}) and
collecting coefficients of powers of $x$, we obtain recursion
formulas for calculating the yet undetermined $C_i^k$ and also the
corrections to eigenenergies  $E_k$
\begin{eqnarray}\label{eq19}
&& C_i^k = \frac1{2 C_0^0} \Biggl[ \Theta(2 - 2 k + i) \Biggl( \gamma_k (-1)^i (3 - 2 k + i) + F_{i-2} \delta_{k,2} - \Biggr. \Biggr. \nonumber\\
&& - \Biggl. \frac{m_{i + 2 - 2k}}{m_0} \Biggl( \gamma_k - \frac{1}{r_0} C_{2 k - 2}^{k - 1} - \sum_{j=0}^{k} \sum_{p=0}^{2 k - 2} C_p^j C_{2 k - 2 - p}^{k - j} \Biggr) \Biggr) - \frac{3 - 2 k + i}{r_0} C_i^{k - 1} - \nonumber\\
&& \Biggl. - \sum_{j=1}^{k - 1} \sum_{p=0}^{i} C_p^j C_{i - p}^{k - j} - 2 \sum_{p=1}^{i} C_p^0 C_{i - p}^k  \Biggr] , \quad i \ne 2 k - 2 ,
\end{eqnarray}

\begin{equation}\label{eq20}
E_k = \frac1{2 m_0} \left( \gamma_k + F_0 \delta_{k,2} -
\frac{1}{r_0} C_{2 k - 2}^{k - 1} - \sum_{j=0}^{k} \sum_{p=0}^{2 k
- 2} C_p^j C_{2 k - 2 - p}^{k - j} \right)
\end{equation}
where  $\Theta(k)$ is the Heaviside step-like function
($\Theta(k)=1$ for $k\geq 0$).

The derived formulas completely resolve the task of calculating
the energy spectrum of the PDM Schr\"odinger equation.

As an illustration, let us compute the energy eigenvalues in the
field of the Coulomb potential $V(r)= - q/r$ in the presence of
the PDM
\begin{equation}\label{eq22}
    m(r)=\frac{m_{\mathrm{c}}}{(1 + ar)^{\lambda}}
\end{equation}
where the constants $q$, $m_{\mathrm{c}}$ and $a$ are assumed to
be positive. We choose the paramet\-ers $\Lambda$, $A$ and $B$ in
the form
\begin{equation}\label{eq23}
\Lambda = \hbar ( n_r + l + 1 ), \quad A = -2 n_r -1, \quad B =
n_r^2 + n_r ,
\end{equation}
which in the constant-mass case guarantees that the Balmer formula
for energies is reproduced just in the zeroth approximation $E_0$
of the $\hbar$-expansion method. The ambiguity parameters are set
to be $\alpha=\gamma=0$, $\beta=-1$ in correspondence with the
kinetic part of the effective Hamiltonian obtained for a crystal
with slowly-varying inhomogeneity \cite{young,bendaniel}.

The calculated partial sums $E^{(k)}=E_0+E_1 \hbar+\ldots+E_k
\hbar^k$ of the $\hbar$-expansions for energies as well as the
exact energy values $E_{\mathrm{num}}$ found by numerical
integration are listed for various $\lambda$ in Table 1. The
calculation has been carried out for the states with  $n_r=0$,
$l=2$ and $n_r=1$, $l=1$, using $q = 10$, $a= 0.1$ in units in
which $\hbar=2m_{\mathrm{c}}=1$.

\begin{table}[htb]
\label{t1}

\caption{Absolute values of partial sums  $E^{(k)}$ of
$\hbar$-expansions for energies in the Coulomb potential in the
presence of the PDM $m(r)=m_{\mathrm{c}}/(1 + ar)^{\lambda}$.
$E_{\mathrm{num}}$ is the result of numerical integration. }
\vspace{10pt} \small \centering
\begin{tabular}{|c|l|l|l|l|l|l|l|l|}
\hline \vphantom{\Big(} $k$ &
\multicolumn{4}{c|}{($n_r=0$, $l=2$)}& \multicolumn{4}{c|}{($n_r=1$, $l=1$)}\\
\cline{2-9} \vphantom{\Big(}   & $\lambda=2$ & $\lambda=3$ &
$\lambda=-2$
  & $\lambda=-3$ & $\lambda=2$ & $\lambda=3$ & $\lambda=-2$& $\lambda=-3$\\
\hline
\multicolumn{9}{|c|}{$\phantom{uuuu} |E^{(k)}|$}\\
0 & 1.77778 & 1.18817 & 3.64395 & 4.04566 & 1.77778 & 1.18817 & 3.64395 & 4.04566 \\
1 & 1.94444 & 1.45345 & 3.50153 & 3.83753 & 2.27778 & 1.98401 & 3.21669 & 3.42127 \\
2 & 1.83333 & 1.25876 & 3.58047 & 3.94991 & 2.07556 & 1.66974 & 3.39891 & 3.69143 \\
3 & 1.83000 & 1.24978 & 3.57934 & 3.94732 & 2.05556 & 1.61180 & 3.39064 & 3.67228 \\
4 & 1.83111 & 1.25190 & 3.58015 & 3.94898 & 2.06000 & 1.62647 & 3.39352 & 3.67887 \\
5 & 1.83111 & 1.25184 & 3.58014 & 3.94896 & 2.06000 & 1.62478 & 3.39330 & 3.67837 \\
\multicolumn{9}{|c|}{$\phantom{uuuu} |E_{\mathrm{num}}|$}\\
  & 1.83111 & 1.25183 & 3.58014 & 3.94897 & 2.06000 & 1.62510 & 3.39329 & 3.67834 \\
\hline
\end{tabular}
\end{table}

It should be added that in the particular case of  $\lambda = 2$
the problem has an exact solution \cite{tkachuk} and the
$\hbar$-expansion technique restores it. Indeed, in this case we
have $E_5 = E_6 = \ldots = 0$ and $E^{(4)}$ coincides with the
exact value, as seen from Table 1.

\section{Order of energy levels}
\label{part4}

Let us apply the obtained recursion formulas to determine the
order of energy levels in the Coulomb field in the presence of
PDM.

First of all, we notice that for the chosen values of the
ambiguity parameters $\alpha=\gamma=0$, $\beta=-1$
there exists an inequality
making comparison of the energy levels with their constant-mass
positions. Namely, if $m(r)$ and $m_{\mathrm{c}}$ are the PDM and
the constant-mass respectively, the corresponding energy
eigenvalues $E(n_r,l)$ and $E_{\mathrm{c}}(n_r,l)$ satisfy
\begin{equation}\label{eq200}
E(n_r,l)\gtrless E_{\mathrm{c}}(n_r,l)\quad\mbox{if}\quad\forall
r\in [0, \infty)\quad m(r)\lessgtr m_{\mathrm{c}},
\end{equation}
provided that these eigenvalues exist.

The inequality (\ref{eq200}) holds valid not only for the Coulomb,
but also for an arbitrary attractive potential since it is a
direct consequence of the variational definition of the discrete
spectrum of a Hamiltonian bounded below  \cite{reed}. In
particular, this inequality explains the increase of binding
energy for impurity bound states in quantum dots that was
predicted in \cite{peter}.

To obtain further inequalities, we fix the shape of the PDM
(\ref{eq22}) and consider the first and second $\hbar$-expansion
corrections whose contribution to the energy is decisive.
According to Eqs.~(\ref{eq10}) and (\ref{eq20}), we have
\begin{eqnarray}\label{eq28}
&& E_0 = - \frac{q}{r_0} + \frac{\Lambda^2}{2 m_{\mathrm{c}} r_0^2} (1 + a r_0)^\lambda , \nonumber\\
&& E_1 = \frac{(2 n_r + 1)(1 + a r_0)^{\lambda} (\omega r_0 -
\Lambda)}{2 m_{\mathrm{c}} r_0^2}
\end{eqnarray}
where
\begin{equation}\label{eq28a}
\omega = \sqrt{\frac{m_{\mathrm{c}} q (1 + a r_0)^{- \lambda - 1}
[a^2 r_0^2 (2 - \lambda)(1- \lambda) + 2 a r_0 (2 - \lambda) +2
]}{r_0 (2 + (2 - \lambda) a r_0)}}.
\end{equation}

It is convenient to rewrite the sum of these two corrections in
the form
\begin{equation}\label{eq29}
E_0+\hbar E_1 = E_{\mathrm{B}}+\frac{m_{\mathrm{c}} q^2}{2
s^2\Lambda^2 } \left[ t^{\lambda}-1+(s-1)^2+\frac{\hbar (2n_r+1)
(\omega r_0/\Lambda-1)}{\Lambda} t^{\lambda} \right]
\end{equation}
with $E_{\mathrm{B}} = - m_{\mathrm{c}} q^2/2 \hbar^2 (n_r + l +
1)^2$ being the constant-mass energy, $t = 1 + a r_0$, $s =
t^{\lambda} - \frac{\lambda}2 (t^{\lambda} - t^{\lambda - 1})$.

Using the expression (\ref{eq29}), we can establish the relative
order of energy levels that refer to the same value of the
principal quantum number $n=n_r+l+1$ and are degenerate in the
constant-mass case. The quantities  $\Lambda$, $r_0$, $\omega$,
$t$ and $s$, appeared in Eq.~(\ref{eq29}), depend on $n$, but not
on $n_r$ and $l$ separately, so they cannot remove the degeneracy.
It is however removed by the last term in the braces due to the
factor $(2n_r + 1)$. The sign of this term coincides with that of
the quantity $(\omega r_0/\Lambda-1)$ for which we find
\begin{equation}\label{eq30}
\frac{\omega r_0}{\Lambda} -
1 = \frac{\omega^2 r_0^2 / \Lambda^2 - 1}{\omega r_0 / \Lambda +
1} = - \frac{\lambda a r_0 \left[ 2 + (3 - \lambda) a r_0 \right]
}{2 (\omega r_0 / \Lambda + 1) (1 + a r_0)^2} .
\end{equation}

Obviously, if  $\lambda < 0$, then $(\omega r_0/\Lambda-1)
> 0 $ and, as a consequence, the level with the quantum numbers $(n_r , l)$
has the higher energy than the level with $(n_r-1, l+1)$.

Now let us prove that for $\lambda > 0$ the relative order of
those levels is converse, i.e., $E(n_r, l) < E(n_r-1, l+1)$. As
seen from Eq.~(\ref{eq30}), it is sufficient to demonstrate that
one has $2+(3-\lambda)ar_0 > 0$ when $\lambda>0$. Rewriting the
equation (\ref{eq9}) for $r_0$ as
\begin{equation}\label{eq31}
r_0 = \frac{\Lambda^2}{m_{\mathrm{c}} q} \left[ t^{\lambda} -
\frac{\lambda}2 (t^{\lambda} - t^{\lambda - 1}) \right]
 = \frac{\Lambda^2}{m_{\mathrm{c}} q} s ,
\end{equation}
we see that it has a positive root only if $s
> 0$. This entails $ar_0 = t - 1 < 2 / (\lambda - 2)$ for
$\lambda > 0$ and, hence, $2 + (3 - \lambda)ar_0 > ar_0
> 0$ that completes the proof.

Thus the inclusion of the PDM
$m(r)=m_{\mathrm{c}}/(1 + ar)^{\lambda}$ removes the
"accidental"\, degeneracy in the Coulomb spectrum so that
\begin{equation}\label{eq32}
E(n_r, l) \lessgtr E(n_r-1, l+1) \quad \mbox{if}
\quad\lambda \gtrless 0.
\end{equation}
Interestingly, in the case of $\lambda < 0$ the above order of the
energy levels coincides with that observed in muonic atoms and
also in the soft-core Coulomb potential \cite{softcore}.

To proceed, let us consider spacings between the levels with the
same value of $n=n_r+l+1$ and evaluate the ratio
\begin{equation}\label{eq33}
R=\frac{E(n_r-1, l+1) - E(n_r, l)}{E(n_r, l) - E(n_r+1, l-1)}.
\end{equation}

We adopt the expression for the energy up to the third correction,
i.e., $E= E_0 + E_1 \hbar + E_2 \hbar^2$. Here $E_2$ calculated
according to Eq.~(\ref{eq20}) reads
\begin{eqnarray}\label{eq34}
&& E_2 = \frac{t^{\lambda - 2}}{16 m_{\mathrm{c}} \omega^2 r_0^4} (n_r^2 + n_r)
\left[-8 \Lambda^2 (a r_0  (\lambda - 2) - 2)^2 - 8 \Lambda \omega r_0
\Bigl( 6 (1 + a_1) - \Bigr. \right. \nonumber \\
&&  \left. 3 a^2 r_0^2  (\lambda - 2) (1 + \lambda + a_1) + a r_0  (12 (1 + a_1)
- \lambda (3 a_1 - 4)) \right) - \nonumber \\
&&\omega^2 r_0^2 \left(a^2 r_0^2  (16 \lambda^2 + 15 a_1^2 + 8 \lambda (3 a_1 + 1)
- 12 a_2 - 8) + 2 a r_0 \left( 15 a_1^2 + 12 \lambda a_1 - \right. \right. \nonumber \\
&& \left. \left. \left. 12 a_2 - 8 \right) + 15 a_1^2 - 12 a_2 -8
\right) \right] + f(n) \equiv  \frac{t^{\lambda - 2}}{16
m_{\mathrm{c}} \omega^2 r_0^4} (n_r^2 + n_r) g + f(n)
\end{eqnarray}
where $f(n)$ is a function that depends on $n$, but not on $n_r$
and $l$ separately. Its explicit form is inessential for us since
we consider the levels with the same $n$.

Combining Eqs.~(\ref{eq29}) and (\ref{eq34}), we find the
requisite ratio
\begin{equation}\label{eq35}
R=\frac{1+(2n_r-1)b_1 / b_2}{1+(2n_r+1)b_1 / b_2}
\end{equation}
where
\begin{eqnarray}\label{eq36}
&& b_1 =\frac{t^{\lambda - 2} \hbar^2 g}{16 m_{\mathrm{c}} \omega^2 r_0^4} , \\
&& b_2 = \left[ \frac{m_{\mathrm{c}} q^2
t^{\lambda}\hbar}{\Lambda^3 s^2} \Bigl(\frac{\omega
r_0}{\Lambda}-1\Bigr) + \frac{t^{\lambda - 2} \hbar^2 g}{16
m_{\mathrm{c}} \omega^2 r_0^4} \right].
\end{eqnarray}

Clearly, one has $R<1$ for $b_1 / b_2 > 0$, $R>1$ for $b_1 / b_2 <
0$. We managed to deduce the sign of $b_1 / b_2$ analytically in
the case of the slowly-varying PDM. Assuming $a$ is small and
performing expansions in its powers, we get $b_1 / b_2 =
\hbar/(\hbar-2\Lambda)+O(a)$. Then from Eq.~(\ref{eq23}) it
follows that $b_1 / b_2 < 0$.

Thus we conjecture the inequality
\begin{equation}\label{eq37}
\frac{E(n_r – 1, l + 1) - E(n_r, l)}{E(n_r, l) - E(n_r + 1, l
-1)}>1.
\end{equation}
It correctness was checked by an exact calculation of the energy
via numerical integration for various, not exclusively small,
values of $a$ and $\lambda$. This inequality says that the shift
on the energy scale is greater for levels with larger orbital
quantum number.

It is worthy to add that a similar problem, the PDM Dirac equation
with the Coulomb potential, was considered in
\cite{VakarchukKepler,VakarchukHeis}. There it was shown that
relativistic factors such as an additional spin-orbit coupling
due to the PDM also affect the order of energy levels.

\section*{Conclusion}
\label{part5}

In the work we have developed the technique for the approximate
calculation of the energy eigenvalues of the radial PDM
Schr\"odinger equation. Based on the expansions in powers of the
Planck constant with the subsequent analysis of the system near
the effective-potential minimum, this technique provides, in
principle, the calculation of the energy up to an arbitrary order
in the analytical or numerical form.

As an example, the spectrum of the Coulomb potential in the
presence of the PDM $m(r)=m_{\mathrm{c}} /(1+ar)^{\lambda}$ has
been investigated and the inequalities regulating the order of the
energy levels in this system have been established. In particular,
the inequalities (\ref{eq32}) and (\ref{eq37}) show how the
"accidental"\, degeneracy of the Coulomb spectrum is removed due
to the PDM.

The expressions for energies obtained with the developed technique
can also be applied to study the dependence of the spectra of
spherically symmetric nano\-structures on their radiuses that is
an actual problem of modern nanophysics.

\vskip 10pt \small { \centerline{
----------------------------------------------------------- }



\begin{thebibliography}{10}

\bibitem{slater} J. C. Slater, Phys. Rev. \textbf{76}, 1592 (1949);
H. M. James, Phys. Rev. \textbf{76}, 1611 (1949); J. M. Luttinger
and W. Kohn, Phys. Rev. \textbf{97}, 869 (1955).

\bibitem{serot} B. D. Serot, J. D. Walecka, Adv. Nucl. Phys. \textbf{16}, 1 (1986).

\bibitem{quantliquid} F. Arias de Saavedra, J. Boronat, A. Polls, A.
Fabrocini, Phys. Rev. B \textbf{50}, 4248 (1994).

\bibitem{clusters} A. Puente, L. Serra, M. Casas, Z. Phys. D \textbf{31}, 283 (1995).

\bibitem{bastard} G. Bastard, \emph{Wave Mechanics Applied to Semiconductor
Heterostructures} (Editions de Physique, Les Ulis, 1988).


\bibitem{young}  K. Young, Phys. Rev. B \textbf{39}, 13434 (1989).

\bibitem{herling}  G. L. Herling, M. L. Rustgi, J. Appl. Phys. \textbf{71}, 796 (1992).

\bibitem{borovitskaya}  E. Borovitskaya, M. S. Shur, Solid-State
Electronics \textbf{44}, 1609 (2000).


\bibitem{li} Y. Li, O. Voskoboynikov, C. P. Lee, S. M. Sze, Solid State
Commun. \textbf{120}, 79 (2001).

\bibitem{voss} H. Voss, J. Comput. Phys.
\textbf{217}, 824 (2006).

\bibitem{muljarov} E. A. Muljarov, E. A. Zhukov, V. S. Dneprovskii,
Y. Masumoto, Phys. Rev. B \textbf{62}, 7420 (2000).

\bibitem{tkach07} M. Tkach, O. Makhanets, A. Hryshchuk,
J. Phys. Stud. \textbf{11}, 220 (2007).

\bibitem{tkach} M. Tkach, V. Holovatsky, O. Makhanets, M. Dovganiuk,
J. Phys. Stud. \textbf{13}, 4706 (2009).


\bibitem{alhaidari} A. D. Alhaidari,  Phys. Rev. A \textbf{66},
042116 (2002).

\bibitem{tkachuk} C. Quesne, V. M. Tkachuk, J. Phys. A: Math. Gen. \textbf{37}, 4267 (2004).

\bibitem{tkachuk2005} B. Bagchi, A. Banerjee, C. Quesne, V. M. Tkachuk, J. Phys. A: Math. Gen.
\textbf{38}, 2929 (2005).




\bibitem{mlodinow} L. Mlodinow,  M. Shatz, J. Math. Phys. \textbf{25}, 943 (1984).

\bibitem{imbo} T. Imbo, A. Pagnamenta, U. Sukhatme, Phys. Rev. D \textbf{29}, 1669 (1984).

\bibitem{weinberg} V. M. Vainberg, V. D. Mur, V. S. Popov, A. V. Sergeev,
A. V. Shcheblykin, Teor. Mat. Fiz. \textbf{74}, 399 (1988).

\bibitem{tutik} S. S. Stepanov, R. S. Tutik,  Teor. Mat. Fiz. \textbf{90}, 208 (1992).

\bibitem{high1N} I. O. Vakarchuk, J. Phys. Stud. \textbf{6},
46 (2002).



\bibitem{roos} O. von Roos,  Phys. Rev. B \textbf{27}, 7547 (1983).

\bibitem{zwaan} A. Zwaan, \textit{Intensit\"{a}ten in Ca-Funkenspectrum}: Thesis (Utrecht, 1929).

\bibitem{dunham} J. L. Dunham, Phys. Rev. \textbf{41}, 713 (1932).

\bibitem{bendaniel} D. J. BenDaniel, C. B. Duke, Phys. Rev. \textbf{152}, 683
(1966).





\bibitem{reed} M. Reed, B. Simon, \emph{Methods of Modern Mathematical
Physics IV: Analysis of Operators} (Academic, New York, 1978).

\bibitem{peter} A. John Peter, K. Navaneethakrishnan, Physica E \textbf{40},
 2747 (2008).

\bibitem{softcore} R. L. Hall, N. Saad, K. D. Sen, H. Ciftci,
Phys. Rev. A \textbf{80}, 032507 (2009).


\bibitem{VakarchukKepler} I. O. Vakarchuk, J. Phys. A: Math. Gen. \textbf{38}, 4727 (2005).

\bibitem{VakarchukHeis} I. O. Vakarchuk, J. Phys. A: Math. Gen. \textbf{38}, 7567 (2005).\\[4mm]


\end{thebibliography}
\end{document}